\documentclass[runningheads]{llncs}
\usepackage{hyperref}
\usepackage{graphicx}
\usepackage{xcolor}
\usepackage{subcaption}
\usepackage{tabularx}
\usepackage{amsmath} 
\graphicspath{{images}}

\usepackage{biblatex}
\usepackage[misc,geometry]{ifsym}

\usepackage[group-separator={,}]{siunitx}
\usepackage[acronym, shortcuts]{glossaries-extra}

\newcommand{\etal}{{\it et al.}}
\glssetcategoryattribute{acronym}{nohyper}{true}
\setabbreviationstyle[acronym]{long-short}

\newacronym{ai}{AI}{Artificial Intelligence}
\newacronym{dl}{DL}{Deep Learning}
\newacronym{gdpr}{GDPR}{General Data Protection Regulation}
\newacronym{ml}{ML}{Machine Learning}
\newacronym{cnn}{CNN}{Convolutional Neural Network}
\newacronym{svm}{SVM}{Support Vector Machine}
\newacronym{pca}{PCA}{Principal Component Analysis}
\newacronym{tab}{TAB}{Text Anonymization Benchmark}
\newacronym{ner}{NER}{Named Entity Recognition}
\newacronym{nlp}{NLP}{Natural Language Processing}
\newacronym{asca}{ASCA}{Acoustic Side-Channel Attack}
\newacronym{sca}{SCA}{Side-Channel Attack}
\newacronym{hid}{HID}{Human Interface Device}
\newacronym{mfcc}{MFCC}{Mel-Frequency Cepstral Coefficient}
\newacronym{tdoa}{TDoA}{Time Difference of Arrival}
\newacronym{voip}{VoIP}{Voice-over-IP}
\newacronym{ofdm}{OFDM}{Orthogonal Frequency-Division Multiplexing}
\newacronym{ms}{ms}{milliseconds}
\newacronym{mm}{mm}{millimeters}
\newacronym{cm}{cm}{centimeters}
\newacronym{hz}{Hz}{hertz}
\newacronym{khz}{kHz}{kilohertz}
\newacronym{svr}{SVR}{Support Vector Regression}
\newacronym{rf}{RF}{Random Forest}
\newacronym{wav}{WAV}{Waveform Audio File Format}

\newcommand{\rev}[1]{\textcolor{blue}}

\usepackage{enumitem}
\newlist{questions}{enumerate}{2}
\setlist[questions,1]{label=RQ\arabic*.,ref=RQ\arabic*}
\setlist[questions,2]{label=(\alph*),ref=\thequestionsi(\alph*)}

\begin{document}
\title{\sf Acoustic Side-Channel Attacks on a Computer Mouse}

\author{Mauro~Conti\inst{1,2} \and
Marin~Duroyon \inst{2}  \and
Gabriele~Orazi\inst{1,4} \and
Gene~Tsudik\inst{3}
}

\institute{University of Padua, Padua, Italy \and
Delft~University~of~Technology, Delft, The Netherlands \and
University of California, Irvine, USA \and
FDM Business Services, Milan, Italy \\ \footnotesize
\email{mauro.conti@unipd.it }
\email{marinduroyon@gmail.com } \\
\email{gabriele.orazi@phd.unipd.it }
\email{gts@ics.uci.edu }
}

\authorrunning{M. Duroyon et al.}

\maketitle              %

\begin{abstract}
\acp{asca}\
extract sensitive information by using audio emitted from a computing devices and their
peripherals. Attacks targeting keyboards are popular and have been explored in the literature.
However, similar attacks targeting other human-interface 
peripherals, such as computer mice, are under-explored.
To this end, this paper considers security leakage via acoustic signals 
emanating from normal mouse usage. 

We first confirm feasibility of such attacks by showing  a proof-of-concept attack that 
classifies four mouse movements with \num{97}\% accuracy in a controlled environment.
We then evolve the attack towards discerning twelve unique mouse movements using a 
smartphone to record the experiment. Using \ac{ml} techniques, the model is 
trained on an experiment with six participants to be generalizable and discern
among twelve movements with \num{94}\% accuracy.  In addition, we experiment
with an attack that detects a user action of closing a full-screen window on 
a laptop. %
Achieving an accuracy of \num{91}\%, this experiment highlights  
exploiting audio leakage from computer mouse movements in a realistic scenario.

\keywords{Cybersecurity \and Human Interface Devices \and Computer Mouse \and Acoustic Signals \and Acoustic Side-channel \and Input Devices \and Audio Leakage\and Machine Learning \and Mouse Movement}
\end{abstract}

\section{Introduction}
\label{sec:intro}

In general, \acp{sca} exploit unintended leaks, such as patterns of power 
consumption~\cite{lipp2021platypus,kogler2023collide}, electromagnetic 
emissions~\cite{yilmaz2019electromagnetic}, visual information (e.g., from screen
content or flickering), or audio (e.g., noise from speakers, CPU or disk) from 
computers or their peripherals to extract sensitive information.
Of those, \acp{asca} typically focus on \acp{hid} -- mechanical devices 
that produce sound when used by humans.
A notable example keyboard acoustic emanations.
Characteristic clicking sounds of a keyboard, generally considered innocuous, 
can be analyzed to infer user keystrokes.
Such attacks occur in the presence of an attacker-placed (or attacker-compromised)
microphone near the target device.
While considerable research has been conducted on keyboard-based attacks~\cite{cecconello2019skype}, 
\acp{asca}~ on mouse movements remain relatively unexplored.

Since computer mice are a very popular means of interaction with laptop and desktop
computers, this work seems to understand potential security vulnerabilities associated with \acp{asca}.
Our objective is to evaluate the correlation between acoustic signals and mouse movements 
thus assessing the feasibility of mouse-based \acp{sca}.
We begin formulating three research questions:
\begin{questions}[itemindent=2em]
    \item \textit{Can mouse sounds leak its motion path and/or activity?}\label{rq1}
    \item \textit{If so, with what accuracy?}\label{rq2}
    \item \textit{In which real-world scenarios do mouse-based \acp{asca} pose a security risk?}\label{rq3}
\end{questions}
This paper is organized as follows:
Section \ref{sec:related_work} explores related work. Next,
Section \ref{sec:methodology} details our research methodology for the experiments.
Then, Section \ref{sec:acoustic_analysis_mouse_move} describes the core experiments,
followed by Section \ref{chapter:real_world}, which discusses security implications of the 
\ac{asca}. Section \ref{chapter:conclusion} concludes~the paper and outlines 
some directions for future works.

\section{Related Work}
\label{sec:related_work}
\acp{sca}~exploit unintended information leakage from computing devices or their peripherals.
For instance, a pattern power consumption of a USB-powered device can be used to profile its 
activity~\cite{spolaor2023plug}. \ac{sca} can even be used to infer the structure 
of a neural network~\cite{chabanne2021side}. This section overviews \acp{asca} --
the category of \ac{sca} that subsumes our work.

Passive \acp{asca}~ operate by interpreting audio emanations. The present work falls into this category 
since the equipment we use (microphone) captures audio without any interactions with the target computer, 
its peripherals or the user.

Gupta \etal\ \cite{gupta2016deciphering} use audio \ac{tdoa}~at two different points (using two microphones 
on the same device) to infer the location of the sound source. Similar studies~\cite{zhu2014context,liu2015snooping} 
explore various methods to recover keystrokes from acoustic emanations. Balagani \etal\ \cite{balagani2022we}
present a novel \ac{asca} on ATM PIN entry, called the PinDrop attack. It involves two steps: (1) 
an acoustic profile is created for each key on the target PIN pad, and (2) the attacker records audio emitted 
by each pressed key during PIN entry and compares these recordings to the acoustic profiles to identify the 
keys pressed. The resulting~classifier is tested on a dataset comprising ten samples for all 
\num{26} English alphabet keys, achieving \num{90.61}\% accuracy.

Cecconello \etal~\cite{cecconello2019skype} present an attack on popular \ac{voip}~software (Skype) 
that captures and transmits all acoustic emanations, including keyboard sounds.
This attack can lead to leakage of sensitive information, such as passwords, that a victim user
might type during a Skype call. Given some knowledge of the victim's typing style and the 
keyboard model, an attacker can achieve a top-5 accuracy of \num{91.7}\% in learning
a random key pressed by the victim. We note that~\cite{cecconello2019skype} is relevant to 
our work since the keyboard is typically used in tandem with the mouse and our 
attack scenario is comparable to that in~\cite{cecconello2019skype}.

PoKeMon~\cite{fang2018eavesdrop} is a new keystroke monitoring method for smartphones using \ac{mfcc}
~\cite{zeng2019spectrogram}. 
Nandakumar \etal\ \cite{nandakumar2016fingerio} propose an innovative finger tracking technique by 
transforming a smartphone into an active sonar system. The system transmits inaudible sounds in 
the \num{18}--\num{20} \ac{khz}~range and tracks the echoes of the finger with its microphones.
The system achieves~2-D finger tracking accuracy of \num{8}~\ac{mm}~at \num{169} frames/sec with a
smartphone prototype.

Cheng \etal\ \cite{cheng2020sonarsnoop} use device speakers to emit inaudible acoustic signals,
which are then reflected off the user's fingers and recorded by the smartphone microphones.
Similarly, \ac{ofdm}~\cite{edfors1996introduction} sounds emitted from device's speakers, 
while the microphones on the same device are then used to record the echos of these sounds.
Also, \cite{cheng2020sonarsnoop} demonstrates that the number of unlock patterns an 
attacker must try until a successful authentication can be reduced by up to 
\num{70}\% using this \ac{asca}.

The two results most relevant to our work are Synesthesia \cite{genkin2019synesthesia} and 
Behavicker \cite{chen4019830behavicker}. 
In the former, Genkin \etal\ examine how \acp{asca}~ can leak screen content via audio emanations,  discovering 
that LCD screens emit content-dependent audio signals which can be captured by nearby microphones.
Experiments conducted using both built-in and webcam microphones reveal acoustic leakage
from audio recordings and video-conferencing, e.g., Google Hangouts.
The authors determine that the power consumption of the monitor's circuits changes depending on the screen content, causing internal components to vibrate and emit sounds.
The authors iteratively simulate the acoustic leakage during hundreds of key presses on the on-screen keyboard, which lead to a \num{100}\% accuracy rate in detecting key presses.
In their attempt at text extraction they achieve an accuracy rate between \num{88}\% to \num{98}\% for most individual characters.
When extracting words from the screen, the correct word is identified in the top five most probable words in \num{72}\% of the cases.

The goal of Behavicker (Chen \etal\ \cite{chen4019830behavicker}) is to determine user activities from 
keyboard and mouse clicking and scrolling events. Sounds produced by keyboard typing or mouse clicking reveal
information about users' behavior. Actions, such as browsing a news website or playing a video game, produce
distinct keyboard and mouse usage patterns, each with its own unique acoustic footprint.
Behavicker identifies six basic interaction events with a \num{88.3}\% accuracy 
and differentiates between seven computer-usage activities with a \num{82.7}\%  accuracy.
It utilizes two functional modules: Acoustic-based Interaction Event Recognition and Computer-Usage 
Recognition. The first uses signal processing and \ac{ml}~to recognize interaction events, 
while the second employs hierarchical classifiers via time-series analysis to distinguish between 
types of activities.

\paragraph{\underline{Research Gap:}}
Our review of related work reveals an appreciable research gap with respect to the analysis of 
audio leakage from mouse activities.
However, mouse usage remains widespread and, similar to keyboards, mice produce a litany of sounds.
Therefore, security of mouse movements deserves and needs to be studied.
This is the key motivation for our work.

\section{Methodology}\label{sec:methodology}
Our work involves multiple phases based on research questions posed in Section~\ref{sec:intro}.
The initial phase is an investigation of mouse-based acoustic leakage.
The second phase is the refinement of our predictive models.
The third and final phase validates these models in real-world scenarios by recording and analyzing audio
from mouse activity in more complex and realistic settings, such as an external mouse used on a laptop.

We do not assume that all mouse activities reflect typical user behavior. Moreover, since this research is
exploratory, the datasets of samples consisted of one participant.
Since collecting these datasets take several days and do not always provide fruitful results, we conducted initial tests with a single sample size. Thus, while we initially ignored plausibility of mouse movements and sample sizes, we considered this later on  in experiments exploring real-world security risks.

We emphasize that all experiments aim to infer the direction of the current mouse movement.
This does not yield pixel-level precision, partly because doing so would require taking 
into account the resolution of the monitor, the model and sensitivity of the mouse as well 
as many other environmental factors. 

All \ac{ml}~models are~trained on a Dell laptop with Intel(R) Core(TM) i7-12700H 2.30 GHz, and 
NVIDIA GeForce RTX 3050, with 16GB of RAM.

\subsection{Mouse and Mouse Pad}
In this study, the \textit{`mouse pad'} is the mouse movement area, either an exterior mouse pad or the flat surface near the laptop,
upon which the latter rests. Initially we experimented with various mouse pads, discovering that, if audio signals are~``audible"
by the microphone it yields the same experimental results.
Similarly, the experimental mouse is a HP X500 mouse, which is a typical commodity office-style model.
For the sake of consistency, the same mouse and mouse pad are used for all experiments.

We use a coordinate system on the mouse pad to categorize mouse movements. Directional movements are defined 
relative to specific points on the pad: `T' for top, `B' for bottom, `L' for left, and `R' for right.
Also, `M' represents the middle along the y-axis, while `C' indicates the center along the x-axis.
Therefore, a movement labelled `TL $\rightarrow$~BR' means a diagonal trajectory from the top-left 
to the bottom-right corner of the pad.

\subsection{Audio Recording Methods}
\label{sec:methodology_audio_recording}

Each experiment uses~different recording methods based on the objective.
Key parameters of audio recordings, such as sample rate, bit depth, channels, and file format, 
play a crucial role in recorded data. Sample rate, typically measured in \ac{khz}, defines the 
number of samples captured per second, where higher rates allow for a more accurate 
representation of the original sound \cite{bits_sample_rates_fundamentals_digital_audio}.
Moreover, according to \cite{bits_sample_rates_fundamentals_digital_audio}, bit depth determines 
the resolution of the amplitude of each audio sample. This influences the noise level of the recording.
The number of channels affects the spatial representation of the sound: mono contains only one waveform, 
while stereo contains two.

Our experiments use~the \ac{wav}, due to being uncompressed, thus ensuring high fidelity of recordings~\cite{bits_sample_rates_fundamentals_digital_audio}.
We record in mono for experiments utilizing a single microphone since
the sound comes from a single source, switching to a stereo configuration in scenarios 
where a second microphone is used. Bit depth selected for each experiment varies from 
\num{16} to \num{32} bits. Sample rate ranges from \num{44.1} \ac{khz}~to \num{48} 
\ac{khz}~in the smartphone range.

In most of our experiments, the microphone is directly connected to the computer. 
To record with a second microphone, using a smartphone, we use an Android application `AudioRec' 
(available in the Google 
PlayStore\footnote{\url{https://play.google.com/store/apps/details?id=com.audioRec&hl=en_US&gl=US}}).

\section{Acoustic Analysis of Mouse Movements}
\label{sec:acoustic_analysis_mouse_move}
Recall that our goal is to find a correlation between audio signals and mouse movements.
While existing literature studied the inference of clicking and scrolling events via 
acoustic signals, the potential for inferring mouse movements from these emissions 
remains unexplored. This section addresses the central research question: 
\begin{quote}
\textit{Can the sounds generated by the movements of a computer mouse be used to infer its trajectory}?
\end{quote}
We initially conduct~experiments using a single microphone, focusing on identifying and 
categorizing distinct movement patterns. Next, we integrate a second microphone to 
further expand the potential of the attack while maintaining a viable attack model.

\subsection{Single Microphone Analysis}
\label{subsec:single_microphone_simple_classification}
\subsubsection{Audio Capture via Processing Software}
\label{subsec:audio_capture_via_processing_software}
\begin{figure}[ht]
    \centering
        \includegraphics[width=0.63\textwidth]{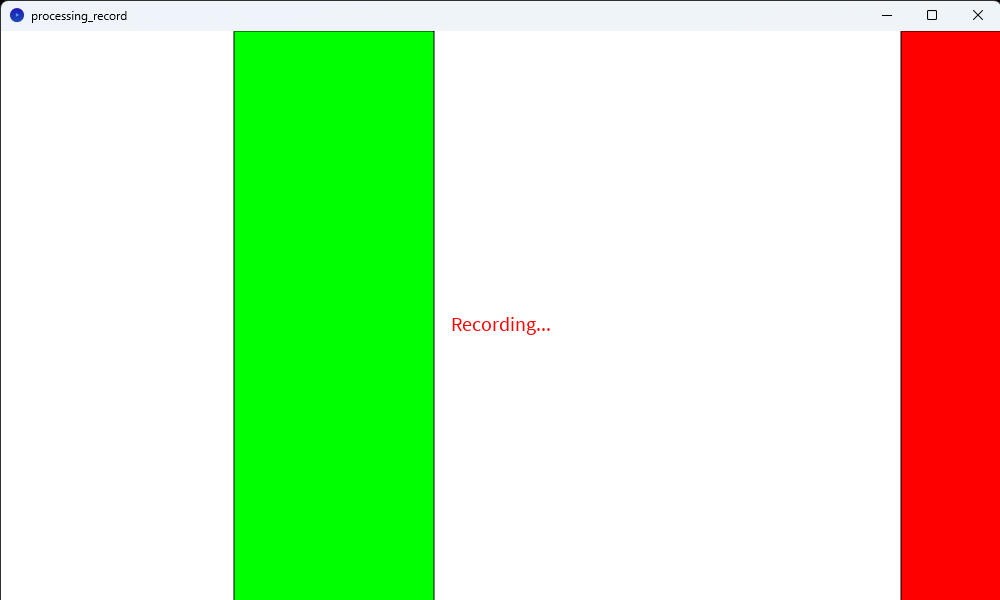}
    \caption{Recording phase of Left to Right movement. The user has to move the cursor from the green area to the red one.}
    \label{fig:processing_record}
\end{figure}

In the first phase, we assess whether an acoustic leakage model could distinguish 
between four basic directional movements of a mouse: up, down, left, and right.
Specifically, the experiment is to determine the feasibility of detecting 
mouse movements along four directions using acoustic signals.
A microphone is positioned on the left side of a right-handed mouse pad.
The mouse is then moved repetitively along the `X' and `Y' axes.
The movements are captured using experimental Java Processing Sketch 
software designed to record these events. As seen in Figure \ref{fig:processing_record}, 
the starting point (in green) triggers the beginning of the recording 
process, while the red-colored end-point stops it.
As a result, the Java Processing Sketch returns acoustic representations 
of mouse movements in various directions.

We collected~\num{6000} samples for each direction, yielding a total of \num{24000} samples.
We note that acoustic signal data-frames, due to the characteristics of the recording procedure, 
are not of a uniform size. Samples are~pre-processed and cleaned for a \ac{cnn}~\cite{intro_cnn} 
\ac{ml}~model. We process the waveform of noises to obtain \acp{mfcc}~using 
the \textit{Sliding Windows} method with a size of \num{36} \ac{ms}, which lead to the best result. 
At this point, we end up with two lists: (1) one of \acp{mfcc}~and (2) the other -- of encoded 
labels from 0 to 3 corresponding to each direction. The list is split into training and testing 
samples, with the testing dataset taking up \num{35}\% of the original dataset size.
The data are entered in a \ac{cnn}~model, compiled with the default Adam optimizer and 
categorical cross-entropy as a loss function using ten epochs.

\begin{figure}[ht]
    \centering
    \includegraphics[width=0.85\linewidth]{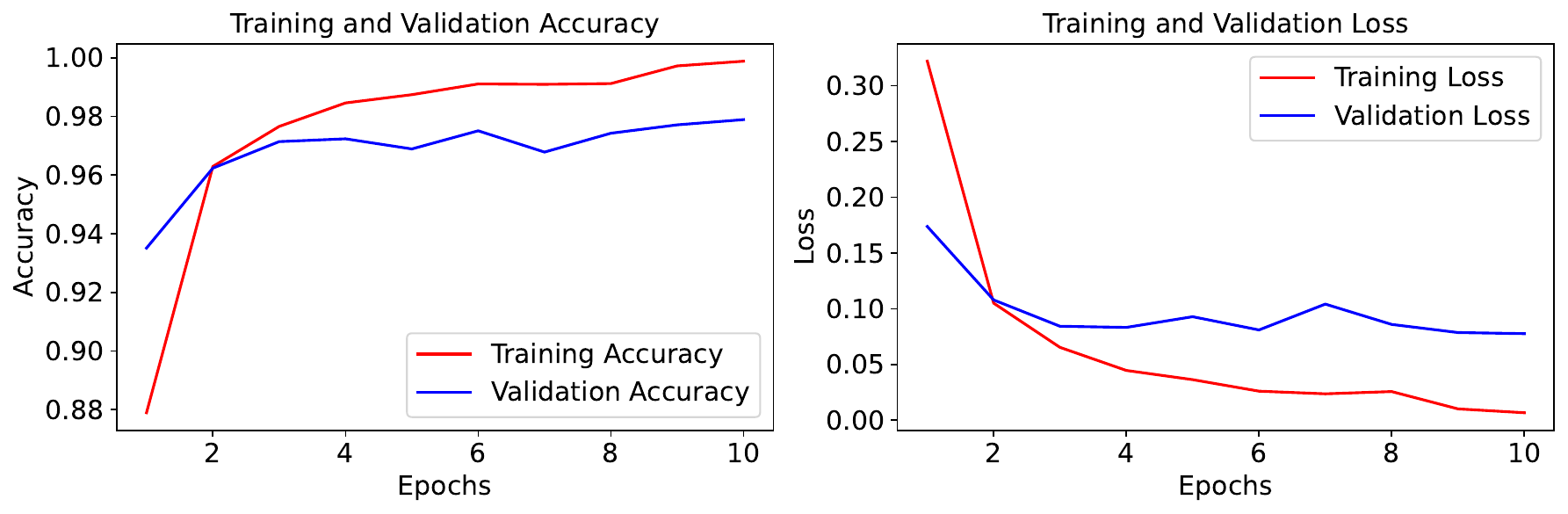}
    \caption{Accuracy and loss graph for classifying four movement categories using Processing Record script.}
    \label{fig:mfcc_4_classes_graphs}
\end{figure}

While this model may appear complex, the experiment shows~no signs of 
over-fitting~\cite{ying2019overview}, as confirmed by the loss functions in 
Figure~\ref{fig:mfcc_4_classes_graphs}. We also verified lack over-fitting with a 
second validation dataset split before any computations. After several checks, we deem the 
results of this experiment as reliable and representative.
This model demonstrates~a high level of confidence, achieving a \num{98}\%  
classification accuracy rate, with F1-scores ranging between \num{97}\% and \num{99}\%
for each category. These performance metrics are reflected
in Table \ref{tbl:mfcc_4_classes_scores}.

\begin{table}[ht]
    \centering
    \begin{tabular}{c|c|c|c|c}
    \hline
    \textbf{Class} & \textbf{Precision} & \textbf{Recall} & \textbf{F1-Score} & \textbf{Support} \\ \hline
    Up & 0.97 & 0.97 & 0.97 & 2093 \\
    Left & 0.99 & 0.99 & 0.99 & 2084 \\
    Down & 0.99 & 0.99 & 0.99 & 2144 \\
    Right & 0.97 & 0.97 & 0.97 & 2079 \\ \hline
    \multicolumn{3}{r|}{\textbf{Accuracy}} & {0.98} & {8400} \\ \hline
    \end{tabular}
    \caption{Classification Report for classifying four movement categories using Processing Record script.}
    \label{tbl:mfcc_4_classes_scores}
\end{table}

This study highlights the viability of \ac{asca}.
While the current experiment is limited due to simple and constrained movements, 
it still serves as a proof-of-concept, motivating further research.

\subsubsection{Continuous Audio Capture During Mouse Movements}
The following experiment we remove the recording software of the previous experiment
and use continuous audio capture corresponding to a more realistic setup.
This means that the mouse moves on the desktop without predefined patterns
and the audio is recorded simultaneously.
In order to set a direction vector and a movement angle corresponding to 
movements in an audio frame, we synchronized the background recording 
with the mouse movements.

The experiment consists~of repeating mouse movements in cardinal directions.
The mouse coordinates are then converted to direction vectors and associated with 
acoustic signal frames. The size of each~acoustic signal chunk is formed by \num{8192} 
samples~with a frame rate of \num{44.1} \ac{khz},  resulting in~approximately 
\num{186} milliseconds. For classification, the angles are categorized into four 
(cardinal) directions using the angle representation of the direction vector,
meaning that \num{25}\% of the circle spatial dimension was associated to a single label.
For instance, with a clockwise circle with 0\textdegree~at the top (north), the angles 
between -\num{45}\textdegree~and \num{45}\textdegree~corresponded to the up (or north) direction category.

\begin{figure}[ht]
    \centering
    \includegraphics[width=0.85\linewidth]{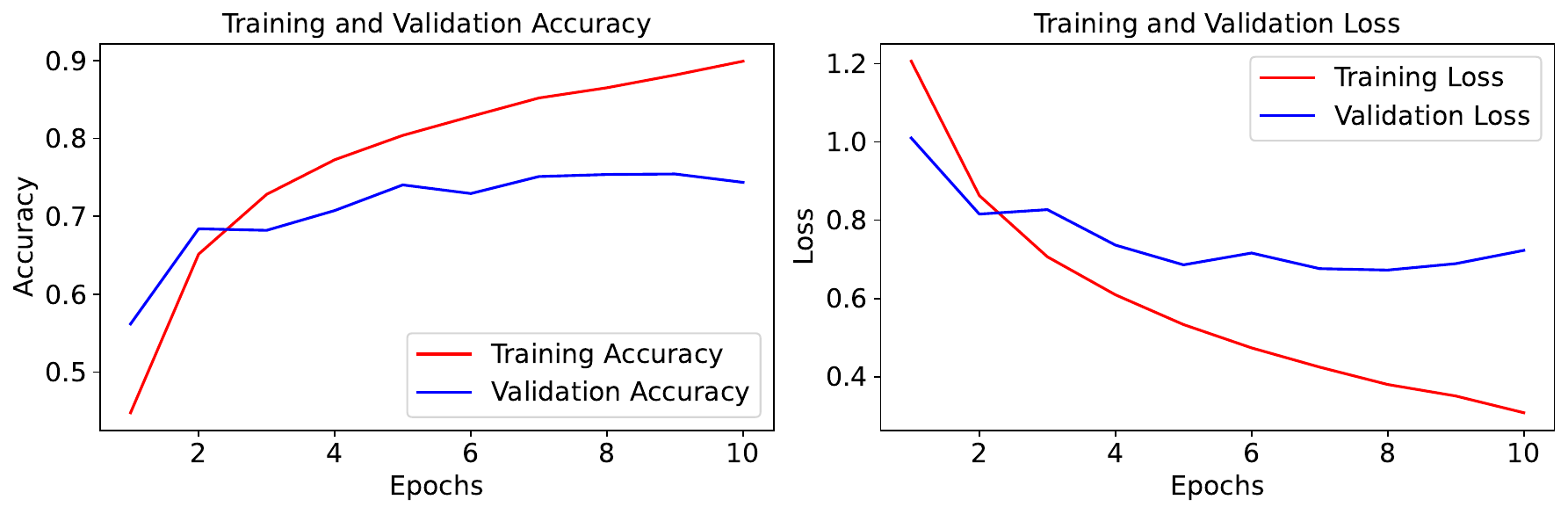}
    \caption{Accuracy and loss graph for four area classification using background recording.}
    \label{fig:background_record_class_graphs}
\end{figure}

Computation of \acp{mfcc}~and their normalization are the only pre-processing steps required  
and are computed using the same methods as in the previous experiment.
Using ten epochs with the same \ac{cnn}~model, we observe a plateau of the validation loss, 
which leads us to assume that the \ac{ml} model stops improving.
Based on the batch size of 32 data points and the \textit{``sum\_over\_batch\_size"} reduction 
method during loss calculation, we obtain a loss value of 0.8, which is higher than 
the average loss for each data point.
With this setting, we achieved the overall accuracy of 74\%. 
The classification report can be seen in Table \ref{tbl:background_record_class_report},
along with the accuracy and loss function graph in Figure \ref{fig:background_record_class_graphs}.

\begin{table}[ht]
    \centering
    \begin{tabular}{c|c|c|c|c}
    \hline
    \textbf{Class} & \textbf{Precision} & \textbf{Recall} & \textbf{F1-Score} & \textbf{Support} \\ \hline
    Up & 0.69 & 0.74 & 0.72 & 755 \\
    Left & 0.80 & 0.68 & 0.73 & 852 \\
    Down & 0.77 & 0.79 & 0.78 & 698 \\
    Right & 0.72 & 0.78 & 0.75 & 765 \\ \hline
    \multicolumn{3}{r|}{\textbf{Accuracy}} & {0.74} & {3070} \\ \hline
    \end{tabular}
    \caption{Classification Report for four area classification using background recording.}
    \label{tbl:background_record_class_report}
\end{table}

The accuracy is lower than that obtained with the previous experiment since the 
recording setup has a greater degree of freedom. With \num{74}\% accuracy we can 
state with confidence that movements in four directions leak enough audio information 
to determine some user activities. Using a pipeline of a Standard Scaler to normalize, 
\ac{pca}~and a \ac{svm}, we obtain the accuracy of \num{83}\%.
Consequently,~this section answers the leakage model research question:
There is indeed an acoustic leakage model for \ac{sca}, and it is 
possible to use audio to infer mouse movements.

\subsubsection{Limitation of single microphone approach}
Previous experiments show acoustic leakage that can be split into four categories, However, 
it does not represent realistic mouse movements.
Of course, the idea is to increase the level of granularity in order to understand what 
angles are discernible with a single microphone and the \ac{ml}~algorithm.

We now evaluate our model to distinguish between eight directions simultaneously, 
doubling the bins of the previous classifier (classes from \num{0} to \num{360}\textdegree~with 
a step of \num{45}\textdegree). The classification report shows that the model has an accuracy 
of at most \num{15}\%.  Despite employing the \ac{ml}~framework similar to prior successful models, 
the algorithm's performance with the eight different categories is sub-optimal.

With different \ac{ml}~models and multiple datasets that cover various environments, 
the results are still falling short: from \ac{svr}~to more complex \acp{cnn}. (These experiments 
are not described due to space limitations.) 

No model could differentiate between eight movement angle categories.
It thus becomes~clear that a singular microphone is~a limiting factor 
in capturing similar mouse movements.
This prompts us to explore other methods.

\begin{figure}[ht]
    \centering
    \begin{subfigure}[b]{0.42\textwidth}
        \includegraphics[width=\linewidth]{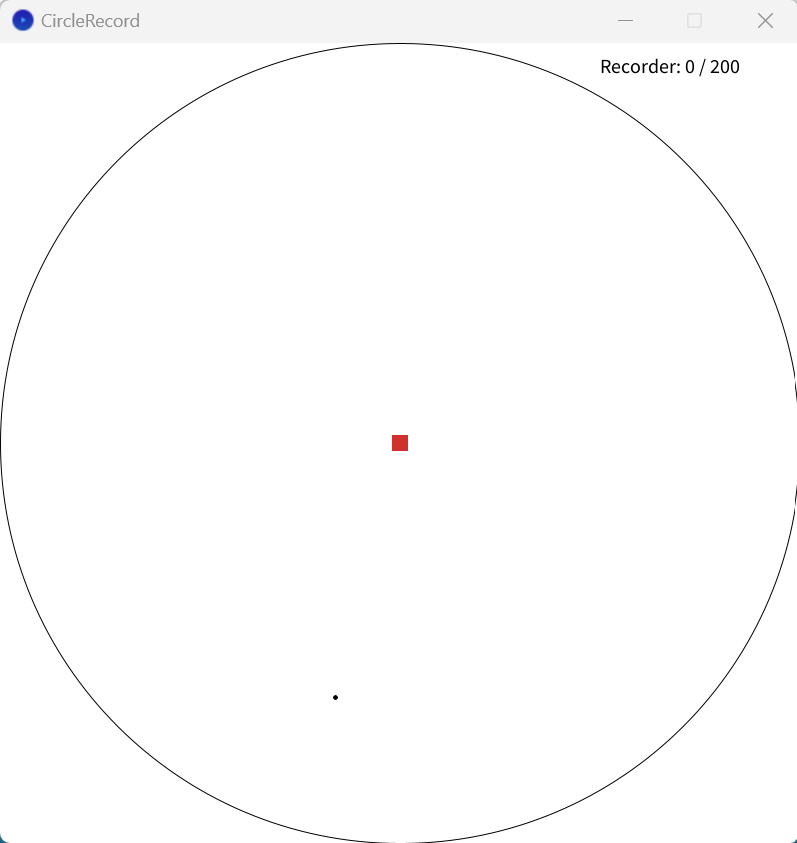}
        \caption{Move cursor towards red center to start the recording.}
        \label{fig:circle_record_redcenter}
    \end{subfigure}
    \hfill %
    \begin{subfigure}[b]{0.42\textwidth}
        \includegraphics[width=\linewidth]{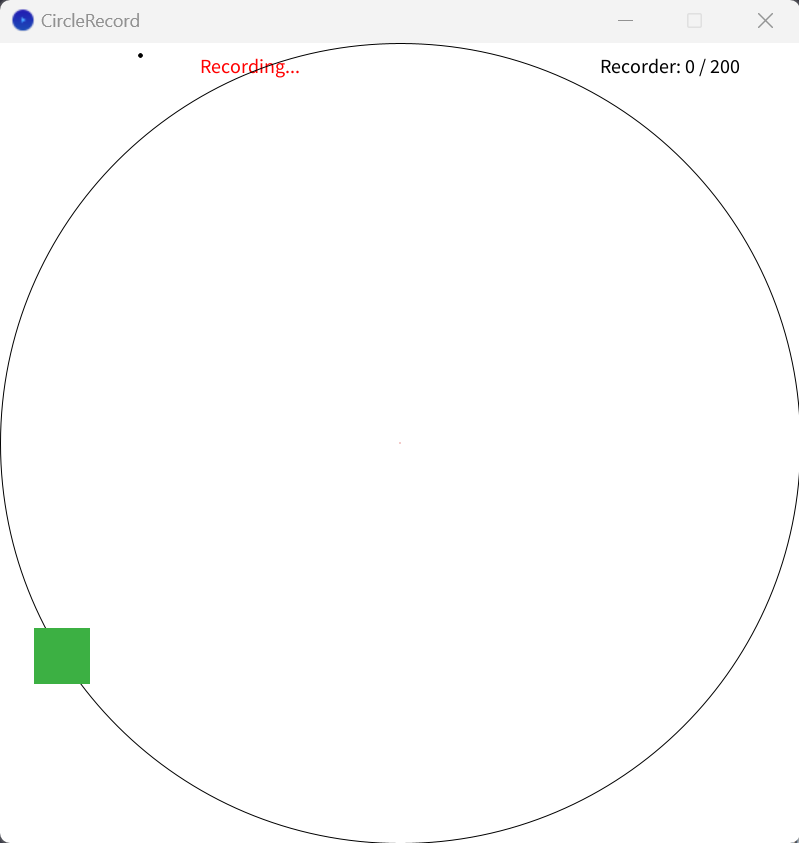}
        \caption{Move cursor towards green box to stop the recording.}
        \label{fig:circle_record_w_target}
    \end{subfigure}

    \caption{Circular recording method.}
    \label{fig:new_circular_recording_method}
\end{figure}

Rather than classifying angles into bins, we attempt to apply a regression learning 
algorithm to predict the angle continuously. The data points are~recorded using a similar 
approach adopted in Section \ref{subsec:audio_capture_via_processing_software}, though 
using a circular pattern. The software, shown in Figure~\ref{fig:new_circular_recording_method}, 
initiates the recording when the mouse reaches the middle of the screen (i.e., a red square) 
and stops it when the green target box is touched by the cursor.  The green boxes are always placed 
on top of the circumference depending on a certain angle with respect to the 
starting point, i.e., the circle center.

The \ac{cnn}~model comprises convolutional layers and fully connected layers to map to the 
target angle prediction. We use a~L1 loss function, which simply calculates the mean 
absolute difference, to approach the target angle as closely as possible.
The dataset is~composed of $\sim$\num{21000} samples of audio \acp{mfcc}~and mouse movement 
angles. The results clearly show randomness in the angle predictions.
Thus, we believe that predicting angles from \num{0} to \num{360} degrees is 
complicated due to the circular nature of angle measurements.
For instance, angles close to \num{0} degrees and \num{360} degrees are nearly 
identical in orientation but are numerically distant.
This discontinuity, also known as ``Angle Periodicity" \cite{li2017deep}, can complicate 
regression models. To address the issue, we conduct~a new experiment where angles are 
represented using their cosine and sine values \cite{li2017deep}.
However, the average angle difference between the model output and the real angle 
is~\(88.79^{\circ}\), which is close to the random angle choice.

To rule out the possibility that the size of the dataset compromises the outcome of the 
regression, we test data augmentation by adding white noise.
However, the model clearly shows~signs of over-fitting on randomized augmented data.

The series of regression experiments conducted using a single microphone yield 
insights into the possibility of accurately predicting continuous mouse movement angles.
However, a single microphone is insufficient to regress the angle of mouse movements.

\subsection{Analyzing Proximity Through Sound Amplitude Variations}
\label{sec:one_microphone_amplitude}
The methodology for analyzing acoustic signals requires~a new approach.
To this end, we examine the amplitude of sound waves to infer the 
proximity of sound-generating movements to a microphone.
Sounds produced nearer the microphone register with higher amplitudes 
than those originating at a greater distance. This variance in amplitude could 
provide a metric for determining the relative distance of the sound source along an axis.

The test involves a controlled series of bidirectional movements along the microphone's 
directional orientation axis with ten samples. During this process, we segment the audio 
recordings into two categories: from left to right to represent the mouse trajectory away 
from the microphone and from right to left to denote its approach towards the microphone.
Next, we plot the audio amplitude over time. To do so, we segment the sound into discrete 
windows and calculate the amplitude, which are then be compared and plotted to visualize 
waveform differences in time. Afterwards, we apply~a linear regression analysis to 
amplitude data points for a better graphical interpretation. This facilitates 
a general trend which reveals whether there is a consistent increase or decrease in amplitude.
The slope of the regression line serves~as the indicator of this relationship.

To highlight the outcomes, we incorporate three distinct auditory markers, 
consisting of three circles done with the mouse at the start and at 
the end of the recording. This modification aims to extend the time frame 
in which the microphone captures the audio.

\begin{figure}[ht]
    \centering
    \begin{subfigure}{\textwidth}
        \centering
        \includegraphics[width=0.9\linewidth]{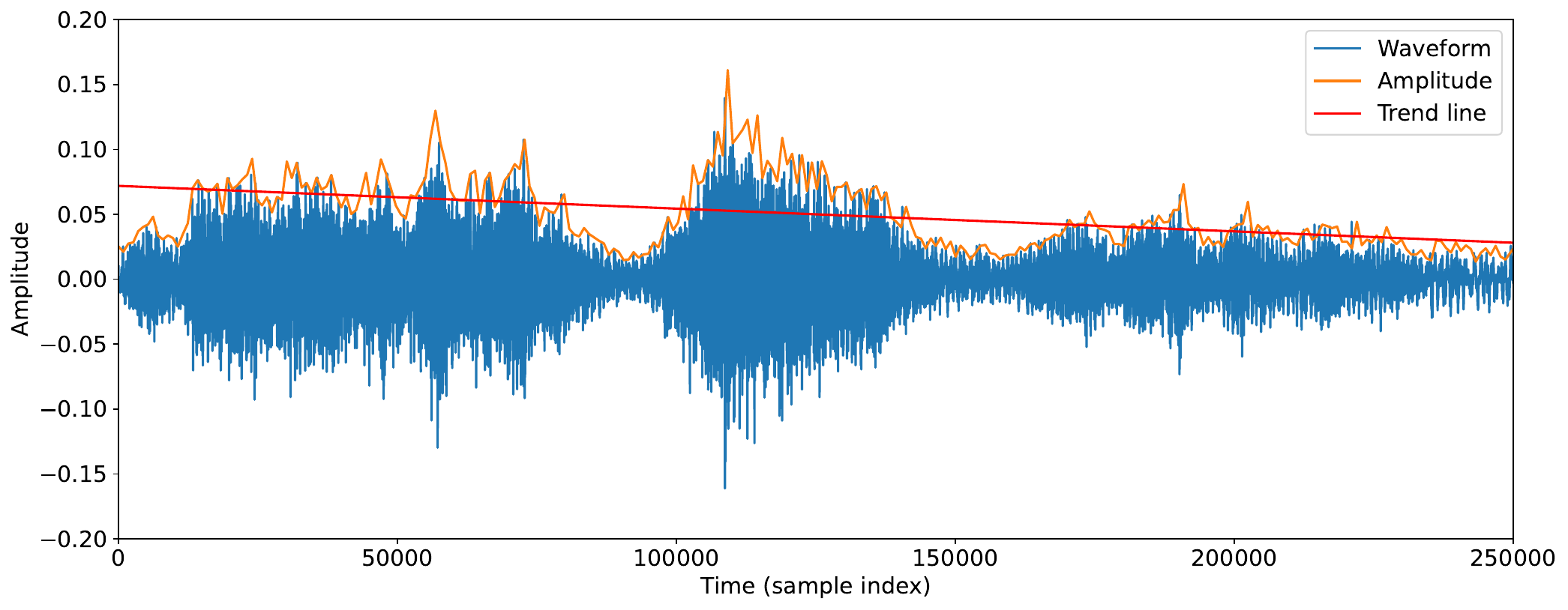}
        \caption{Waveform of the sound moving away from the microphone (descending amplitude).}
        \label{fig:arabian_jasmin_away}
    \end{subfigure}
    \hfill
    \begin{subfigure}{\textwidth}
        \centering
        \includegraphics[width=0.9\linewidth]{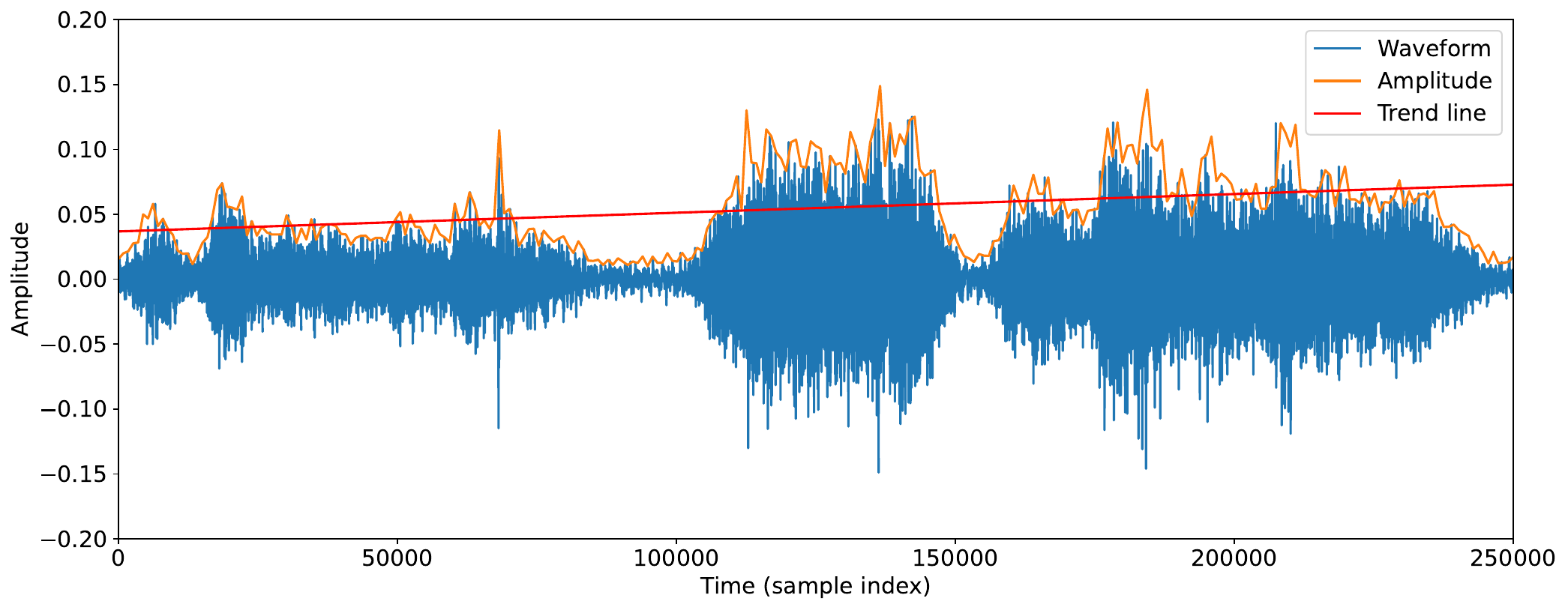}
        \caption{Waveform of the sound moving towards the microphone (ascending amplitude).}
        \label{fig:arabian_jasmin_towards}
    \end{subfigure}
    \caption{Audio waveforms for the mouse moving towards and away from the microphone.}
\end{figure}

The experimental data in figures \ref{fig:arabian_jasmin_away} and \ref{fig:arabian_jasmin_towards} 
illustrate the subtle, yet consistent, trends in sound amplitude corresponding to the mouse movement 
near the microphone.
We obtain a negative slope angle value of \( -0.0000099^{\circ}\) decrement in amplitude 
when the mouse moves away from the microphone, and a positive slope angle value of 
\( 0.0000068^{\circ}\) in the opposite direction.
Although minimal, the differences are consistent across different tests.
These outcomes demonstrate the possibility to visually determine the direction of a mouse movement, validating the hypothesis that a distinctive mouse movement along a single axis can be 
accurately inferred.

\subsection{Dual-Microphone Approach for Two-Dimensional Amplitude Analysis}
\label{sec:dual_mic_amplitude_analysis}
\begin{figure}[ht]
  \centering
  \includegraphics[width=0.55\textwidth]{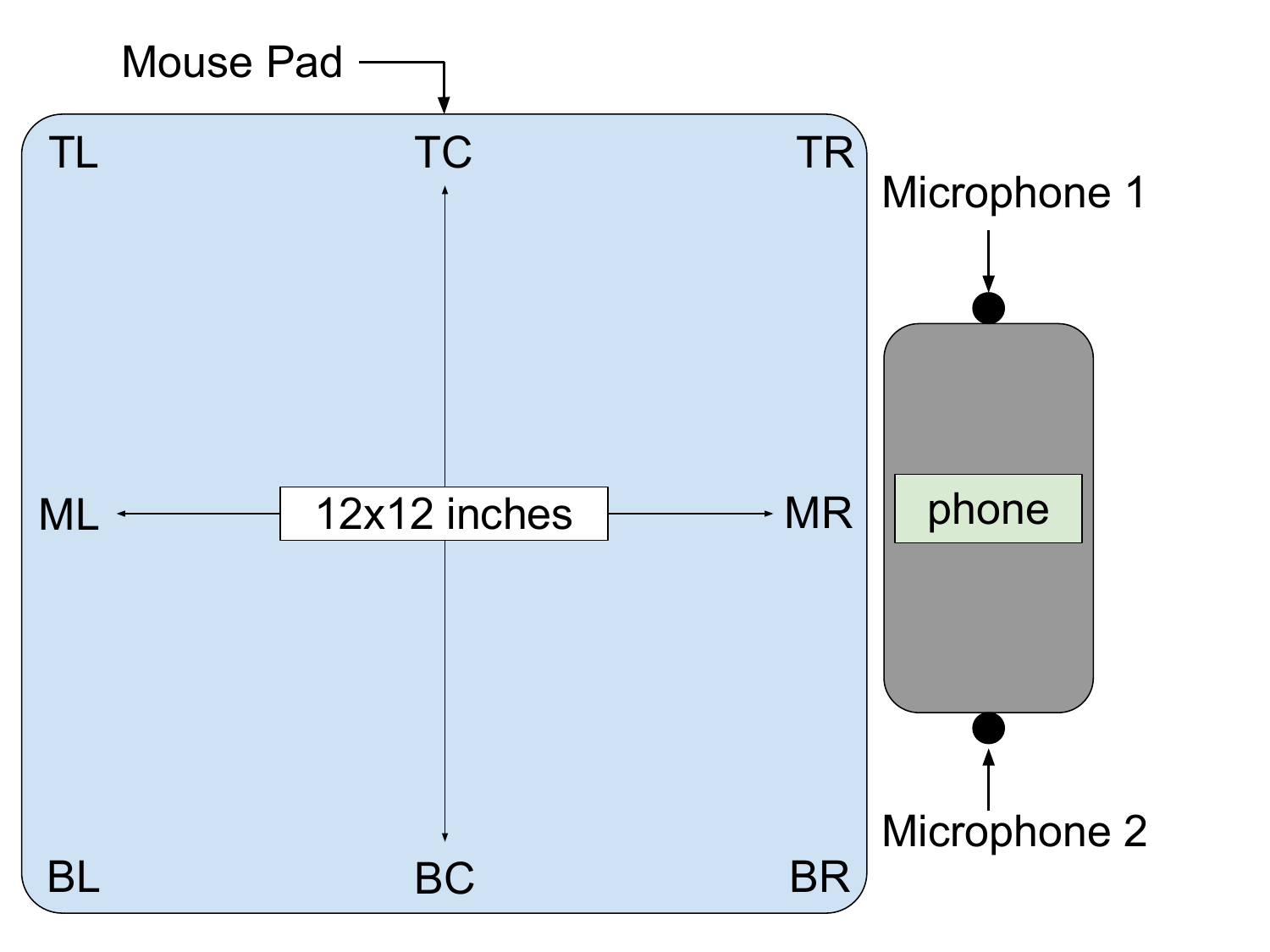}
  \caption{Dual-Microphone experimental setup.}
  \label{fig:seals_british_setup}
\end{figure}

We now introduce a second microphone. This is achieved using two microphones of a 
smartphone to provide a two-dimensional measurement plane, as shown 
in Figure \ref{fig:seals_british_setup}.

\begin{figure}[h]
    \centering
    \begin{subfigure}{\textwidth}
        \centering
        \includegraphics[width=0.9\linewidth]{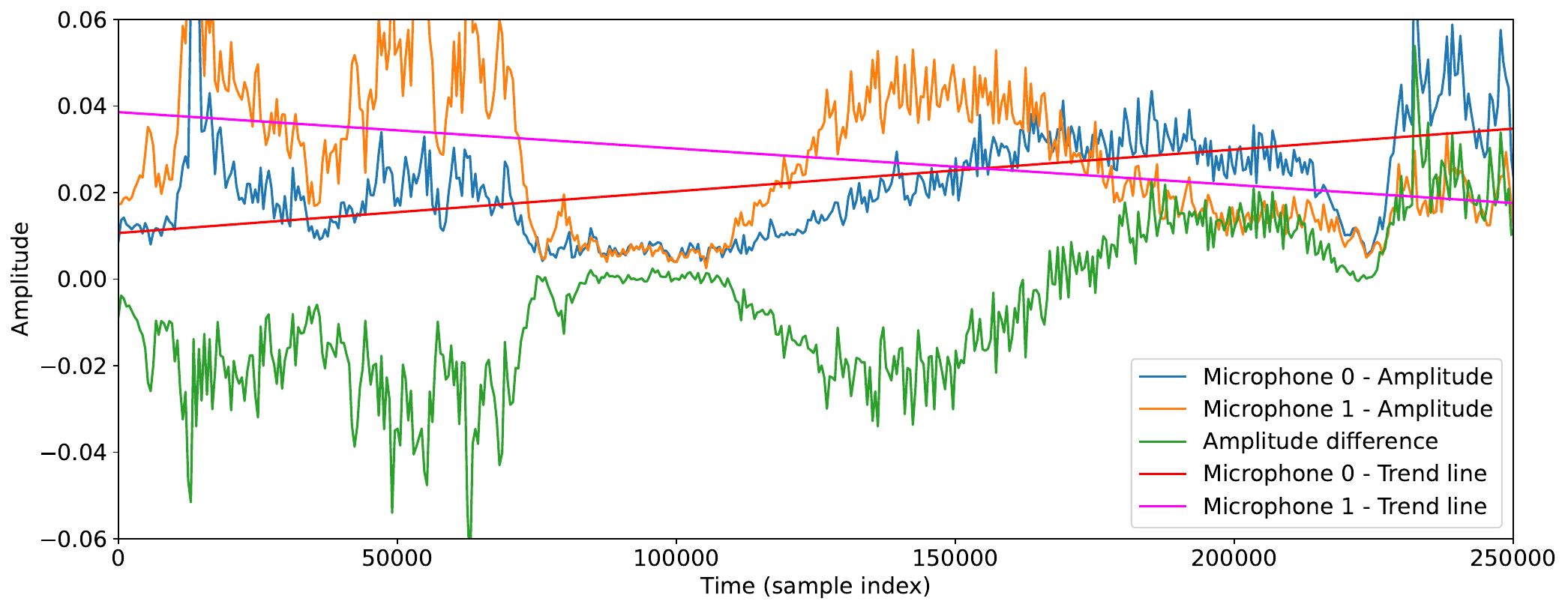}
        \caption{Amplitude difference and trend when moving vertically up the table.}
        \label{fig:seals_british_02.pdf}
    \end{subfigure}
    
    \begin{subfigure}{\textwidth}
        \centering
        \includegraphics[width=0.9\linewidth]{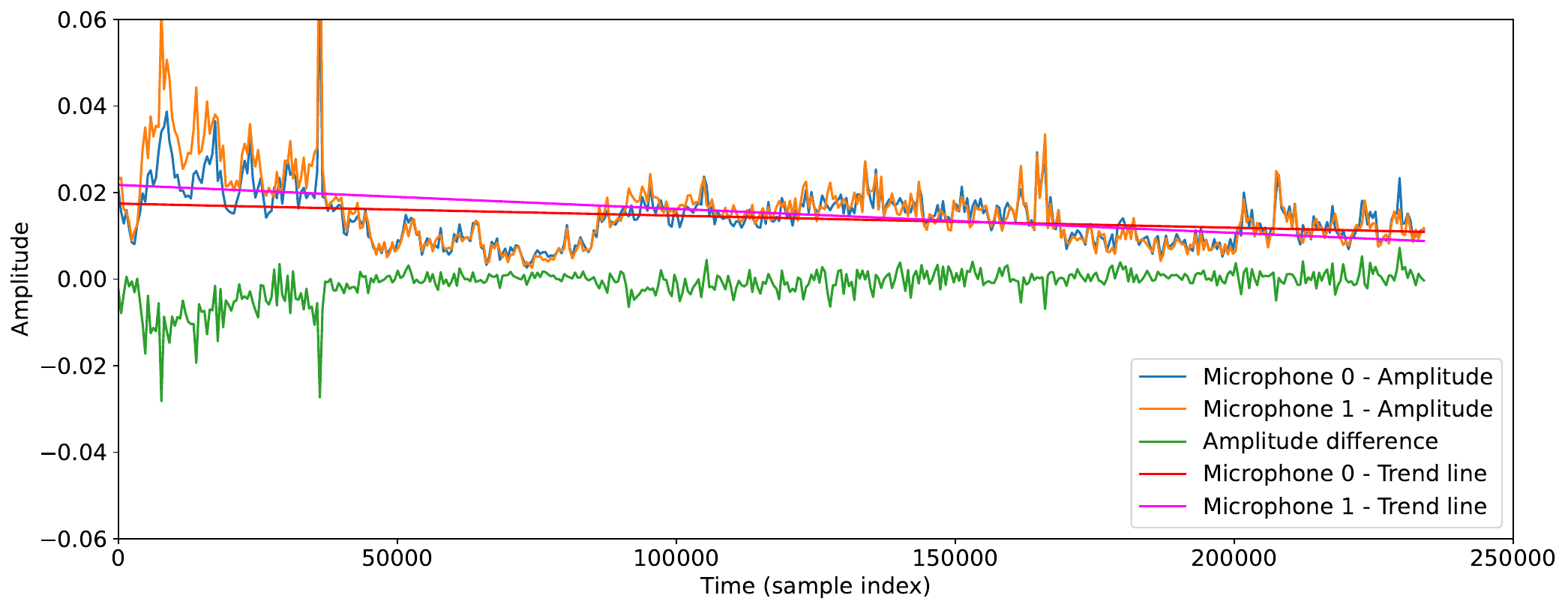}
        \caption{Amplitude trend with a negative slope, indicating movement away from the phone.}
        \label{fig:seals_british_03.pdf}
    \end{subfigure}
    
    \caption{Graphs showing amplitude differences and trends based on the mouse's movement.}
    \label{fig:seals_british_graphs}
\end{figure}

The recording movement pattern involves~traversing the length of the mouse pad, 
moving from top to bottom and back, and then horizontally from a location
near the phone to the far edge and back. Figure \ref{fig:seals_british_graphs} shows the 
resulting graph of two out of four cardinal directions.
Each graph shows~a plot of the two microphone amplitudes and the amplitude difference.
Also, regression lines for the microphone amplitudes are plotted to show general trends.
As can be seen in Figure \ref{fig:seals_british_02.pdf}, representing moves up the mouse pad, 
a criss-cross effect is evident, indicating one microphone ``hearing" louder than the 
other one at certain times.
Furthermore, Figure \ref{fig:seals_british_03.pdf} shows a negative slope trend in the 
regression line when moving away from the microphone. Even though graphs are not included in 
Figure \ref{fig:seals_british_graphs} for the sake of simplicity, we observed coherent 
behaviour while moving the mouse towards or down with respect to the smartphone.

Since the experiment yield visual representations that clearly distinguish various 
directional movements, we extend it by attempting to distinguish among ten movement patterns.
We record and segment approximately \num{250} samples for each directional shift 
using Audacity\footnote{\url{https://www.audacityteam.org/}}. The ten categories are a 
combination of movements with specific start and ending points of the mouse pad, 
where each point is defined by a vertical and horizontal location.
These points are shown in Figure \ref{fig:seals_british_setup}.

\begin{table}[ht]
    \centering
    \begin{tabular}{c|c|c|c|c}
    \hline
    \textbf{Class} & \textbf{Precision} & \textbf{Recall} & \textbf{F1-Score} & \textbf{Support} \\ \hline
    BC$\rightarrow$TC & 0.91 & 0.94 & 0.92 & 52 \\
    BL$\rightarrow$BR & 1.00 & 1.00 & 1.00 & 50 \\
    BR$\rightarrow$BL & 1.00 & 0.96 & 0.98 & 50 \\
    BR$\rightarrow$TR & 1.00 & 0.97 & 0.99 & 39 \\
    ML$\rightarrow$MR & 0.83 & 0.88 & 0.85 & 49 \\
    MR$\rightarrow$ML & 0.88 & 0.86 & 0.87 & 49 \\
    TC$\rightarrow$BC & 0.98 & 0.98 & 0.98 & 52 \\
    TL$\rightarrow$TR & 0.98 & 1.00 & 0.99 & 49 \\
    TR$\rightarrow$BR & 0.97 & 0.92 & 0.95 & 39 \\
    TR$\rightarrow$TL & 1.00 & 1.00 & 1.00 & 49 \\ \hline
    \multicolumn{3}{r|}{\textbf{Accuracy}} & {0.95} & {478} \\ \hline
    \end{tabular}
    \caption{Classification report on ten movements with dual-microphone setup.}
    \label{tbl:class_report_nicole_test}
\end{table}

Using a \ac{rf}~Classifier with \num{100} estimators, the model achieves a 
high accuracy of \num{95.19}\%, indicating good performance across various movement categories.
The classification report in Table \ref{tbl:class_report_nicole_test} provides further 
insights into the model's efficacy.

In summary, this experiment demonstrates~the ability to infer ten distinct two-dimensional 
movements, thus exposing the ability to understand the leakage model of a 
mouse using a smartphone.

\section{Real-world Implications and Security Risks}
\label{chapter:real_world}

We now consider practical implications of our results.
A typical attack environment could be a shared office (e.g., a cubicle farm),
a cafeteria, or a library. A simple smartphone represents the attack vector.
Modern smartphones now feature dual microphones, mainly to reduce ambient noise during calls.
The smartphone is a perfect attack vector since it can safely be placed near the victim.
It requires no interaction in the recording phase and can remain with its screen off the entire time, so as 
not to arouse suspicion. With this scenario in mind, we now describe experiments with six participants who 
volunteered to be recorded while performing predefined mouse movements. 
Based on this, we examine real-world implications  of such attacks.

\subsection{Experiment with Other Participants}
The next phase involves extending the experiments to include a multiple participants.
This helps~to determine the generalizability of the model and its applicability to 
various environments. The new experiment involves~six 
participants performing a series of mouse movements on a mouse pad. 
This setup is identical to the one described in Section~\ref{sec:dual_mic_amplitude_analysis}.

\begin{figure}[ht]
    \centering
    \includegraphics[width=0.55\linewidth]{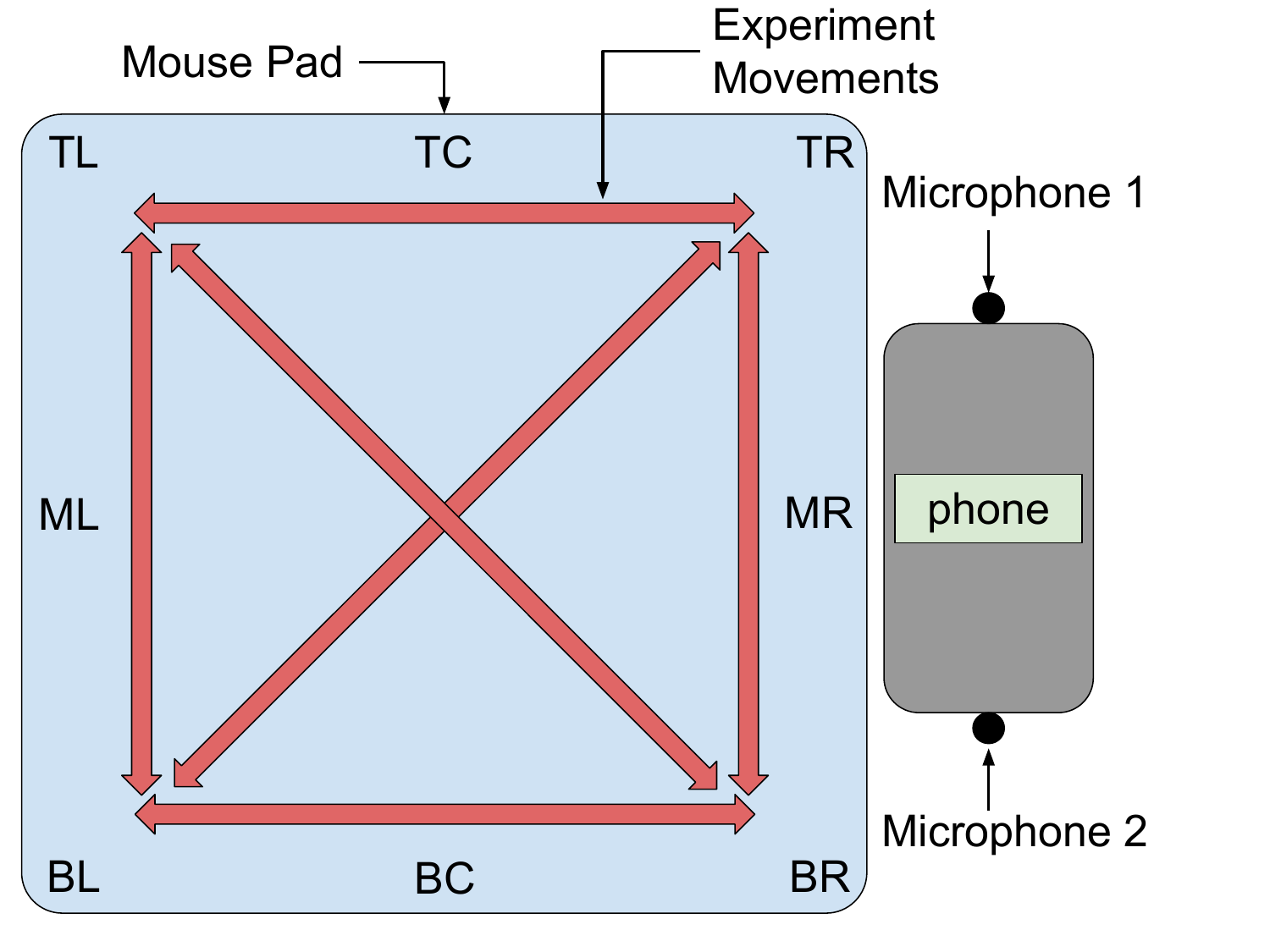}
    \caption{Mouse pad with mouse movements in the experiment.}
    \label{fig:mouse_pad_mov_chap7}
\end{figure}

Participants are~instructed to perform distinct back-and-forth movements along both axes 
within a 12-by-12 mouse pad placed on an office table.
The experiment details~specific movement patterns, shown in Figure \ref{fig:mouse_pad_mov_chap7}.
Participants are asked to repeat each back-and-forth movement for one to two minutes, 
resulting in a total experiment time ranging from \num{6} to \num{12} minutes.
To help with the segmentation during the pre-processing phase, participants are~also 
asked to clap before and after recording each axis to delineate the start and end.

The six participants show variability in their mouse usage methods.
For example, some execute the movements rapidly, resulting in a higher frequency of 
actions, while others move more slowly.
In addition, the amount of pressure applied to the mouse differs~among participants, 
producing distinct acoustic signatures. These speed, frequency, and pressure differences 
are~integral to the experiment's design. They provide a diverse dataset that is~ideal 
for testing the generalizability of the \ac{ml}~model, since it reflects~a more 
realistic range of user behavior.

We specifically isolate the segments corresponding to mouse movements using a thresholding 
technique to distinguish them from background noise, followed by a manual verification 
process to ensure accuracy. The waveforms from both channels are~split into \num{50} 
discrete windows and then we calculate their amplitude. Thus, we extract the same number 
of data points for each movement acoustic signal regardless of different recording methods 
used by the participants.

Our analysis also involves~calculating the difference in amplitudes between the top and the 
bottom microphone recordings for each segment. Consequently, we generate a ``difference amplitude line'' 
of \num{50} data points representing the differential amplitude for each audio sample.
These data points serving as learning values are the `X' values.
The `Y' values, or target labels, are~assigned based on the type of directional movement recorded.
This structured dataset, with its distinct X and Y values, allows~us to train and test the 
\ac{ml}~model to classify the directional movements based on acoustic signatures. 

This experimental setup is~designed to record six types of back-and-forth mouse movements, 
resulting in \num{12} distinct directional classes. In the course of the experiments, a total 
of \num{5507} samples are~collected. The distribution of these samples across the different 
classes is between \num{406} and \num{506} samples each, with a mean of \num{456} samples per class.

As shown in Table \ref{tbl:class_report_multi_participants}, the classification report evaluates 
performance of the \ac{ml}~model. The report indicates high accuracy in the model's predictions, 
with a low rate of false positives. The F1-scores, which combine precision and recall into a single 
measure, consistently reflect high performance, predominantly in the range of \num{0.93} to \num{1.00}.
This underscores the model's balanced accuracy in both precision and recall dimensions.
The overall test accuracy of the model is \num{96}\%, indicating the model's effectiveness 
in accurately classifying the directional movements.

\begin{table}[ht]
\centering
\begin{tabular}{c|c|c|c|c}
\hline
\textbf{Direction} & \textbf{Precision} & \textbf{Recall} & \textbf{F1-Score} & \textbf{Support} \\ \hline
BL$\rightarrow$BR & 0.98 & 0.99 & 0.98 & 81 \\
BL$\rightarrow$TL & 0.91 & 0.98 & 0.95 & 97 \\
BL$\rightarrow$TR & 0.96 & 0.90 & 0.93 & 91 \\
BR$\rightarrow$BL & 0.99 & 0.93 & 0.96 & 82 \\
BR$\rightarrow$TL & 0.94 & 0.99 & 0.96 & 91 \\
BR$\rightarrow$TR & 1.00 & 0.99 & 0.99 & 89 \\
TL$\rightarrow$BL & 0.92 & 0.97 & 0.94 & 96 \\
TL$\rightarrow$BR & 0.98 & 0.93 & 0.96 & 92 \\
TL$\rightarrow$TR & 0.99 & 1.00 & 1.00 & 101 \\
TR$\rightarrow$BL & 0.93 & 0.95 & 0.94 & 91 \\
TR$\rightarrow$BR & 0.98 & 0.93 & 0.95 & 90 \\
TR$\rightarrow$TL & 0.99 & 0.99 & 0.99 & 101 \\ \hline
\multicolumn{3}{r|}{\textbf{Accuracy}} & 0.96 & 1102 \\ \hline
\end{tabular}
\caption{Classification Report for a Random Forest Classifier with \num{100} estimators.}
\label{tbl:class_report_multi_participants}
\end{table}

\subsection{Inferring Realistic Mouse Movements}
\label{sec:clicking_x_exp}
To demonstrate the feasibility of such attacks and help in assessing their potential impact on 
real-world scenarios, we shift towards a more practical approach. The next experiment aims to 
determine whether it is possible to detect when a user clicks the `close' button, typically 
found at the top right of a Windows laptop screen. While closing a window may not directly reveal 
sensitive security information, the ability to discern precise user interactions like button 
clicks might have broader implications for user privacy.

The experiment is~structured around two recording sessions of five minutes with a single participant. 
It is~performed on a traditional mouse setting using a Windows laptop and an entry level office mouse pad.
In the first session, the participants are~instructed to move the cursor from random points on the screen 
to the top-right corner and click the red `X', a standard action for closing a window (positive class for the classification).
The second session involves~recording mouse movements followed by clicks at various random points on 
the screen -- a pattern different from the first recording session (negative class).

We focus on a time window around click events in the analysis phase. This approach allows us to compare
acoustic characteristics of mouse movements and clicking sounds associated with both positive and 
negative classes.

The recordings are first analyzed to determine the clicking events through quantile thresholding methods.
To do so, the waveforms are converted to a mono channel to calculate the absolute value of the waveform.
This threshold can be modified based on the visual representation and manually adjusted.

\begin{figure}[ht]
    \centering
    \includegraphics[width=\linewidth]{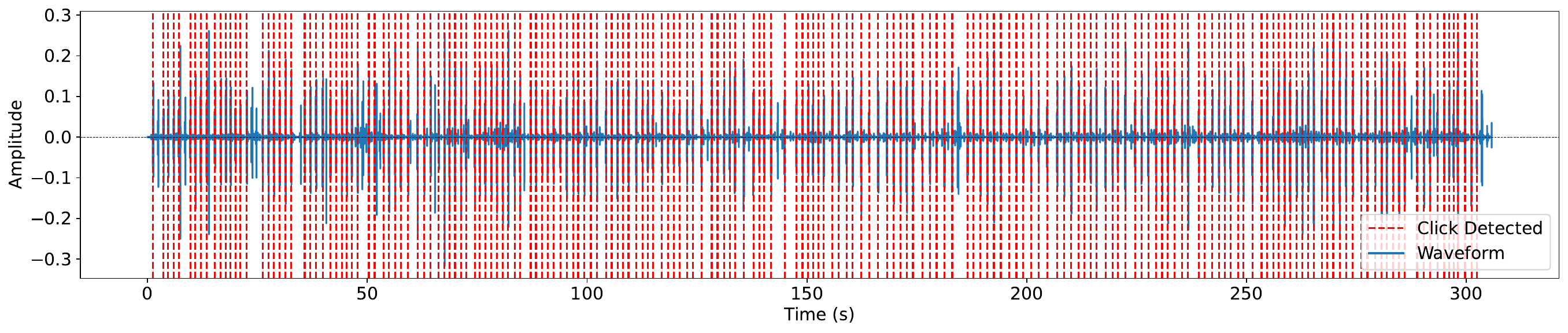}
    \caption{Detected clicks out of mouse events.}
    \label{fig:blue_camo_detecting_clicks}
\end{figure}

With these clicking events, the goal is~to find the two consecutive peaks corresponding 
to the mouse press and release events during a mouse click. To do so, we use~a minimum and 
maximum distances between detected peaks to isolate the events that seem to correspond to mouse-clicking events.
These distances are again estimated via empirical values, in this case, from \num{10}~\ac{ms}~to 
\num{200}~\ac{ms}. Figure~\ref{fig:blue_camo_detecting_clicks} shows the waveform with red 
vertical lines representing detected clicks.

As we iterate through each identified mouse click event, a specific time window surrounding 
each event is extracted to prepare the data for the \ac{ml}~pipeline.
Next, we apply~an \ac{mfcc}~transformation to each window.
These transformed data points are fed into a \ac{ml}~pipeline utilizing a binary 
\ac{rf}~Classifier with \num{50} estimators.

The classification report shown in Table \ref{tbl:classification_report} reflects the performance 
of the binary classification model. The model demonstrates a high level F1-score, with class 
\textit{`General Click'} reaching \num{0.92} and class \textit{`Closing Click'} at \num{0.89}.
These scores indicate a strong balance between precision and recall across both classes.
The overall accuracy of the model stands at \num{91}\% for the \num{136} samples tested, 
further reinforcing the model's effectiveness.

\begin{table}[ht]
    \centering
    \begin{tabular}{c|c|c|c|c}
    \hline
    \textbf{Class} & \textbf{Precision} & \textbf{Recall} & \textbf{F1-Score} & \textbf{Support} \\ \hline
    General Click & 0.89 & 0.96 & 0.92 & 77 \\
    Closing Click & 0.94 & 0.85 & 0.89 & 59 \\ \hline
    \multicolumn{3}{r|}{\textbf{Accuracy}} & {0.91} & {136} \\ \hline
    \end{tabular}
    \caption{Classification Report for the Binary Classification Model.}
    \label{tbl:classification_report}
\end{table}

This experiment successfully demonstrates~the ability to distinguish between various mouse 
movements and their associated clicks. Using controlled recordings, we train a \ac{rf}~Classifier 
to recognize these specific activities. 

The goal of the next phase is to extend this approach to more natural computer usage scenarios.
We record the authors' typical computer activities, including working on a project~in Overleaf, 
ending with the action of closing a window.
These recordings are~then processed using the steps previously described.
Next, we apply the pre-trained \ac{rf} Classifier to the isolated mouse click events 
and infer the type and nature of mouse movements and clicks.
This step marks a significant advance in applying our model to real-world scenarios, thus
bridging the gap between controlled experimental conditions and everyday computer usage.

\begin{figure}[ht]
    \centering
    
    \begin{subfigure}{\textwidth}
        \includegraphics[width=\linewidth]{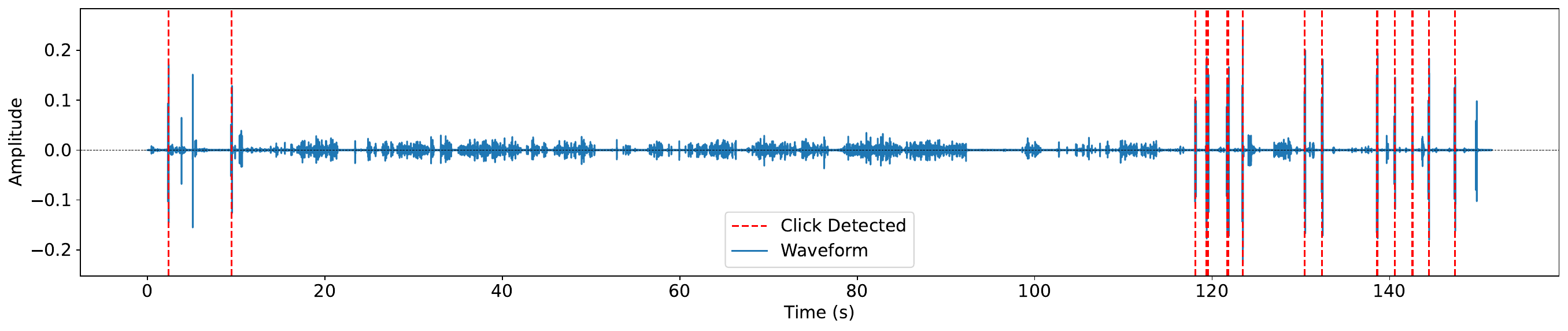}
        \caption{First experiment recording.}
        \label{fig:blue_camo_bc_1}
    \end{subfigure}
    \hfill
    \begin{subfigure}{\textwidth}
        \includegraphics[width=\linewidth]{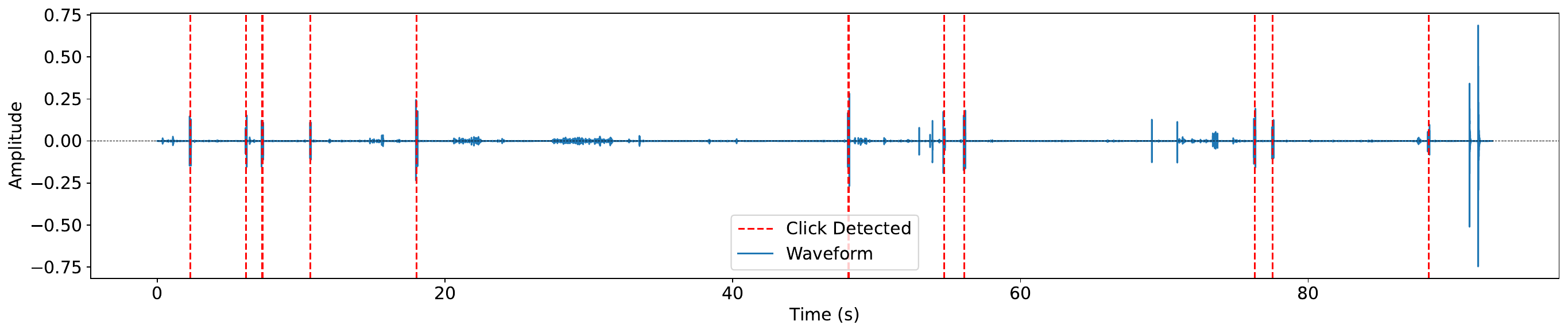}
        \caption{Second experiment recording.}
        \label{fig:blue_camo_bc_2}
    \end{subfigure}

    \label{fig:blue_camo_windows_comp}
    \caption{Comparison of waveform detection in two iterations of the experiment.}
\end{figure}

We experiment by dividing it into two iterations.
In the first iteration, the red `X' is clicked to close the window, together with other two additional clicks.
However, due to the thresholding method employed, the final click is  not detected, as per Figure \ref{fig:blue_camo_bc_1}.
In the second iteration, the red `X' to close the window is  clicked and followed with two hand claps.
This means that the final detected mouse click is the one of interest, as showin in Figure \ref{fig:blue_camo_bc_2}.
Looking at the waveforms, we observe background audio that includes random clicks and 
typing, creating a consistent hum of ambient noise.

In the first recording, among \num{15} clicks detected, the next-to-last sample is  
accurately identified as corresponding to the `clicking red X' event.
Similarly, in the second recording, out of \num{11} notable events, the model correctly infers 
the last one as the `click red X' event. These results demonstrate the trained model's ability 
to accurately identify window-closing events, pointing at the feasibility of employing 
such a method in an \ac{asca}.

\subsection{Security Risks of \acp{asca}~ based on Mouse Movement Inference}
Ourfindings illustrate the feasibility of developing a generalizable model applicable to 
various users and the capability to accurately identify specific mouse movements in realistic settings.
Looking ahead, these insights open the door to potential future threats where \acp{asca}
might be combined with keyboard inference techniques, further reconstructing a victim's computer activities.
For instance, one plausible scenario involves targeting users filling out sensitive online forms, 
such as tax documents. In such cases, the model could be fine-tuned to recognize distinctive 
patterns associated with clicking specific buttons or entering information into text fields.
The attack does not aim to recognize by the number of pixels across which 
the user moved the moue and whether, at any given time, the cursor is precisely over a specific button. 
However, pattern recognition in mouse movements could lead to a comparable result.
This highlights the importance of further research into \acp{asca}.

\section{Ethical Concerns}
\label{subsec:ethconc}
Our institutions do not require any formal IRB approval to perform the experiments described above. 
Nonetheless, all experiments and corresponding evaluations were performed in accordance with 
the guidelines of the Menlo report~\cite{bailey2012menlo}. We preemptively informed all voluntary 
participants about intended usage of their data. We obtained their informed consent before the 
recording process. We anonymously recorded only audio samples produced by a mouse and explicitly 
asked all participants to avoid speaking during recording sessions. We used audio samples 
for research purposes only. 

\section{Conclusions}
\label{chapter:conclusion}
This paper explored mouse-based \acp{asca}. It analyzed whether audio emanations from mouse 
movements reveal any sensitive information. The initial  experiments show that it is possible 
to differentiate among four basic mouse movements. We then looked into granular measurements 
in order to predict the direction angles of a moving mouse, which shows some predictive complexities.

Next, we switched to a stereo recording method using two microphones on a smartphone, This
allowed us to distinguish between ten two-dimensional movements on a mouse pad.

We then considered potential real-world scenarios where mouse-based \acp{asca}~pose a security risk.
To this end, we trained a model -- using six participants -- that predicts \num{12} distinct 
two-dimensional movements. This shows the ability to generalize the model and attack multiple users
in one recording environment.
Furthermore, we experimented with detection of a specific user actions, such as closing a 
Window, which demonstrated the efficacy of our approach in a realistic experimental setup.
Consequently, we showed that the acoustic leakage model of mouse activity
poses a real security risk.

Since this is only the first attempt to experiment with mouse-based \acp{asca}
we identify directions for future work. 
Applying regression algorithms to stereo acoustic data would be interesting, since
all regression tests so far are  performed using one microphone.
Moreover, the experiment in Section \ref{sec:clicking_x_exp} 
could be extended by finding other situations where mouse movements reveal sensitive information.
Finally, mouse-based \acp{asca} can be conducted in tandem with keyboard-based ones.
The combination of the two might yield even more leakage.

\printbibliography[heading=bibintoc,title=References]

\end{document}